# Curriculum Guidance Document

## The BlockNet Consortium

PROJECT BLOCKNET INTELLECTUAL OUTPUT 3
-WHITE PAPER-

# Curriculum Guidance Document

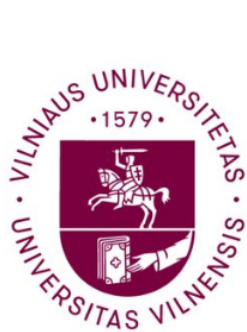 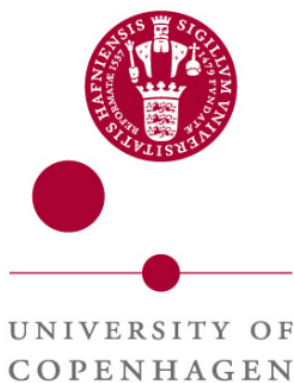 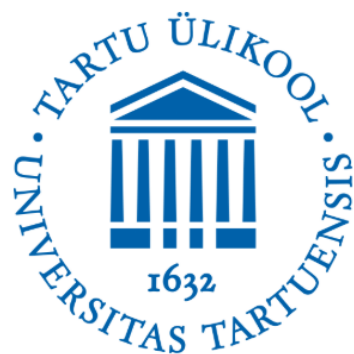

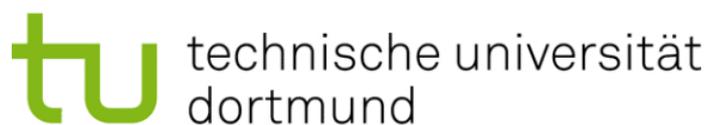

February 2020

Disclaimer
The creation of these resources has been (partially) funded by the ERASMUS+ grant program of the European Union under grant no. 2018-1-LT01-KA203-047044.
Neither the European Commission nor the project's national funding agency DAAD are responsible for the content or liable for any losses or damage resulting of the use of these resources.

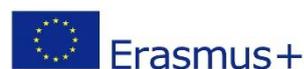

**Project acronym**: BlockNet

**Project name**: BlockChain Network Online Education for interdisciplinary European Competence Transfer

**Output type**: Intellectual Output

**Activity leader**: University of Copenhagen

**Contributors**: TU Dortmund University, Vilnius University, University of Tartu

**Acknowledgement:** The persons of University of Copenhagen involved in preparation of this document part are Boris Düdder and Haiqin Wu from University of Copenhagen. This document has been reviewed and contributed by Michael Henke, Natalia Straub, Tan Guerpinar, Philipp Asterios Ioannidis from TU Dortmund University, Vladislav Fomin from Vilnius University, and Raimundas Matulevičius, and Mubashar Iqbal.

# ABSTRACT

Blockchain is a challenging topic since it is novel and fosters potential innovation. The blockchain is attractive for various disciplines, and, because of its cross-cutting nature, needs knowledge stemming from various disciplines. The devised curriculum can be instantiated specifically to meet the needs of students' groups from various disciplines. The pedagogical innovation of the project is the inclusion of interdisciplinary project groups with participant's interaction via online platforms for project-based learning activities. MOOCs and SNOCs allow blended-learning for interdisciplinary and geographically distributed student groups.

# REQUIREMENTS FOR THE O3

| Requirement # | Requirement summary | Project description |
|---|---|---|
| 1.<br>2. | Defining the (1) **didactical concept** and (2) **methods** for teaching and learning | "The project activity "Identification of didactic and organisational concepts for Blockchain Small Network Online Course" will aim at defining the didactic concept and methods for teaching and learning for higher education study programmes based on blended-mobile learning and problem- and project based learning that allows collaborative work of international students from different institution and studies programs." |
| 3.<br>4.<br>5. | (3) **Module structure**, (4) **exam form** and (5) **learning outcomes** will be defined for each module | "Module structure, exam form and learning outcomes will be defined for each module. Exam form can be oral or written, depending on the lectures and the features of courses." |
| 6. | (6) **Methods to evaluate and measure prior knowledge and learning outcomes** of different strategies will be reviewed | "Existing didactic methods for online learning and various methods to evaluate and measure prior-knowledge and learning outcomes of different strategies will be reviewed." |
| 7. | (7) **Tailored online distance and active learning methods** to develop explicit and tacit competences will be developed and applied | "Tailored online distance and active learning methods to develop not only knowledge but also explicit and tacit competences will be developed and applied." |
| 8. | (8) **Demands, standards and conditions for integration** of the SNOC in existing study programmes | "The organisational concept includes the integration of the SNOC in existing study programmes at the individual institutions. Demands, standards and conditions are covered." |

Disclaimer
The creation of these resources has been (partially) funded by the ERASMUS+ grant program of the European Union under grant no. 2018-1-LT01-KA203-047044.
Neither the European Commission nor the project's national funding agency DAAD are responsible for the content or liable for any losses or damage resulting of the use of these resources.

TABLE OF CONTENTS

Disclaimer
The creation of these resources has been (partially) funded by the ERASMUS+ grant program of the European Union under grant no. 2018-1-LT01-KA203-047044.
Neither the European Commission nor the project's national funding agency DAAD are responsible for the content or liable for any losses or damage resulting of the use of these resources.

# Vocabulary of terms used

Massive Open Online Courses (MOOC) - Offers learning experiences to students, from video lectures, readings, assignments and exams, to opportunities to connect and collaborate with others through threaded discussion forums and other Web 2.0 technologies (Yang, et al., 2013).

Small Network Online Courses (SNOC) – compared to MOOC, a smaller form of an online course with a significantly smaller group of students who potentially are geographically closer located, e.g., in a region, or pre-qualified groups of high-caliber students from select programs within the small network. Offers live classes with a virtual classroom environment.

Constructive Alignment - a principle used for devising teaching and learning activities, and assessment tasks, that directly address the intended learning outcomes in a way not typically achieved in traditional lectures, tutorial classes and examinations (Biggs & Tang, 2011).

Project - teaching format, where students – in contracts to usual lectures – work on tasks utilizing the knowledge they obtained before. Project-based learning allows students to learn by doing and applying ideas. Students engage in real world activities that are similar to the activities that professionals engage in (Krajcik & Blumenfeld, 2005).

Case study - highly adaptable style of teaching that involves problem-based learning and promotes the development of analytical skills. By presenting content in the format of a narrative accompanied by questions and activities that promote group discussion and solving of complex problems, case studies facilitate development of the higher levels of Bloom's taxonomy of cognitive learning; moving beyond recall of knowledge to analysis, evaluation, and application (Bonney, 2015).

Use case - is a methodology used in system analysis to identify, clarify, and organize system requirements. The use case is made up of a set of possible sequences of interactions between systems and users in a environment and related to a particular goal. The method creates a document that describes all the steps taken by a user to complete an activity (Rouse, 2020). In educational context, a use case is used as a "walk through" guidance for students with regard to learning goals and available learning materials.

# Introduction

The didactical and organizational concept for interdisciplinary blockchain SNOC (further in the text "the curriculum concept") is designed for an interdisciplinary audience of higher education students who want to learn about blockchain in interdisciplinary contexts and its specific relevance for their individual knowledge fields. The curriculum concept design defines the demands and requirements, standards, and conditions for the integration of blockchain SNOC in existing study programmes at the individual institutions. It consists of integrated plans for learning, the design of plan implementations, and plan evaluations, their implementations, and assessment of the outcomes of the learning experience. The purpose is framing learning and teaching and translating broad statements of intent into specific plans. The goal is to ensure that students receive an integrated and coherent learning experience (Prideaux, 2003). The curriculum concept design renounces to prescribe ephemeral information because blockchain is a fast-developing topic and avoiding binding a curriculum to outdated information. Additionally, the design respects the knowledge

Disclaimer
The creation of these resources has been (partially) funded by the ERASMUS+ grant program of the European Union under grant no. 2018-1-LT01-KA203-047044.
Neither the European Commission nor the project's national funding agency DAAD are responsible for the content or liable for any losses or damage resulting of the use of these resources.

and skills of course responsible, hence, learning outcomes are not tightly coupled with teaching methods or detailed topics. But we take the liberty to offer suggestions and ideas which are based on recommendations of academia and practitioners.

Blockchain is a challenging topic since it is novel and fosters potential innovation. As a digital technology, blockchain is well suited for online education, and students can benefit from online courses (Barr & Miller, 2013). We propose teaching blockchain using constructive alignment in blended-learning and online courses (Ali, 2018).

# ONLINE TEACHING

## ONLINE COURSES AND PLATFORMS

Different concepts of how courses can be delivered online exist. SNOC stands for "Small Network Online Course" and is a smaller form of a popular concept of online course referred to as MOOC (massively open online course).

To maximize the success of the project and cater for sustainability of knowledge obtained in the past Erasmus+ projects, we build up on the previous work of the PERFECT project (PERFECT Consortium 2018) and its results, which evaluated and devised methods and tools for MOOCs. MOOCs (massively open online courses) bring huge volume of students with varying degrees of interest and capabilities together, and are generally open to anyone, regardless of qualifications. In contrast, a SNOC (small network online course) is a smaller form of an online course with a significantly smaller group of students who potentially are geographically closer located, e.g., in a region, or pre-qualified groups of high-caliber students from select programs within the small network. This term was proposed by the Yale's management school. The initial idea behind SNOC is to provide a means for a group of business schools to offer classes to each other's MBA students within a closed online network.

MOOC typically consists of a mix of self-study materials such as video lectures as well as readings and learner activities such as quizzes, discussions, simulations, and peer-reviewed assignments. The duration of MOOCs varies a lot, generally from 3-10 weeks, with an expected weekly workload of two to eight hours. As the principal difference between SNOC and MOOC is in the number of students, the didactical and organizational concept for SNOC can follow that of MOOC. In practice, as for example, at Coursera (Anon., ) and Moodle (Muhsen, et al., 2013) partially, MOOCs and SNOCs run in sessions. Hence, multiple sessions comprise a module.

A cohort of learners will begin the course simultaneously and go through the course together, week by week. In this way, learners will have classmates to discuss course issues with and peers to review assignments. However, a learner can quickly race ahead of the course schedule since all course materials are available as soon as the learner is enrolled. Doing so might lead to the problem of having to wait a bit for responses to discussion posts and peer-reviews of assignment as well as learner's possibility to review other learners' assignments. If a learner falls behind, he or she can join the next session, and the progress will be transferred. Students draw additional benefits from digital tools in online task-based or project-based education (Skoogh, et al., 2012). **The innovation of SNOC/MOOC didactics in this project is the inclusion of interdisciplinary project groups with participant's interaction via online platforms.**

8Disclaimer
The creation of these resources has been (partially) funded by the ERASMUS+ grant program of the European Union under grant no. 2018-1-LT01-KA203-047044.
Neither the European Commission nor the project's national funding agency DAAD are responsible for the content or liable for any losses or damage resulting of the use of these resources.

## MOOC AND SNOC DEVELOPMENT PROCESS

The development process of a MOOC and a SNOC does not differ in the necessary development activities. The process for both online courses is characterized by extensive preparation, which can be divided up into the following phases as depicted in Figure 1. The individual phases contain the following activities:

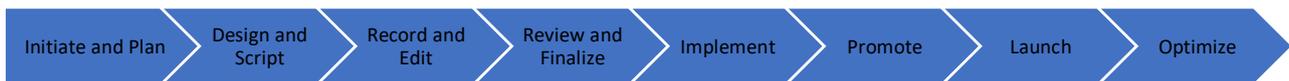

FIGURE 1: EXAMPLE OF A DEVELOPMENT PROCESS OF A MOOC

The process structure is inspired by (McFarlane, 2011) and accommodates for the differences between traditional face-to-face and online teaching.

### INITIATE AND PLAN

The actions and decisions of the first phase build the foundation for an excellent online course that will reach the intended learners. Lecturers have to be engaged in delivering content and teaching the course.

**The definition of the target group and learning objectives are crucial to the course success and definition of the course outline.** It assures that the course content will help the learners to reach the learning objectives and to coordinate content across lectures in a well-organized structure with progression. The creation of the course outline generates ideas for learning activities: quizzes, assignments, and discussions. The additional material types, i.e., ppt slides, reading material, expert interviews, and video lectures should then be selected and designed. **An important point is the planning of peer interactions of learners in a cohort, which includes the evaluation of technical or physical meeting spaces to facilitate peer interactions.**

The next activity is identifying promotion channels to attract potential learners and motivate them to enroll. A short, 1-2 minutes, video trailer for the course, or a visual campaign with images and central questions that will be addressed in the course are attracting learners. The promotion material can be published on relevant websites, distributed on social media and mailing lists.

Creating an online course is a non-trivial task that suggests developing a project plan. The project plan acts as a baseline plan for the course retrospective.

### DESIGN

In the Design & Script phase, lecturers do the central part of their work and develop the course content[1]. The course responsible(s) and a potential production team must ensure a well-structured course through review, coordination, and feedback.

**Instead of creating all course material from scratch, the course responsible should consider using open-access resources that are free, publicly available, online materials, e.g., articles, websites, databases,**

---

[1] Under the BlockNet project, the course content will be developed in the consecutive stage of the Project, as Intellectual Output O4 "Development of Learning Material for Interdisciplinary BlockChain SNOC".

Disclaimer
The creation of these resources has been (partially) funded by the ERASMUS+ grant program of the European Union under grant no. 2018-1-LT01-KA203-047044.
Neither the European Commission nor the project's national funding agency DAAD are responsible for the content or liable for any losses or damage resulting of the use of these resources.

**policies, videos, quiz-sets, and presentations** (e.g., MERLOT https://www.merlot.org/). Referring directly to such materials is linking the course content to real-life challenges and solutions. The later discussed case studies and projects are an excellent source for related material.

Student learning is supported by good visual helping to show key points. The presentation can be built around central visual material such as exciting data visualizations, or a good illustration of causes and effects.

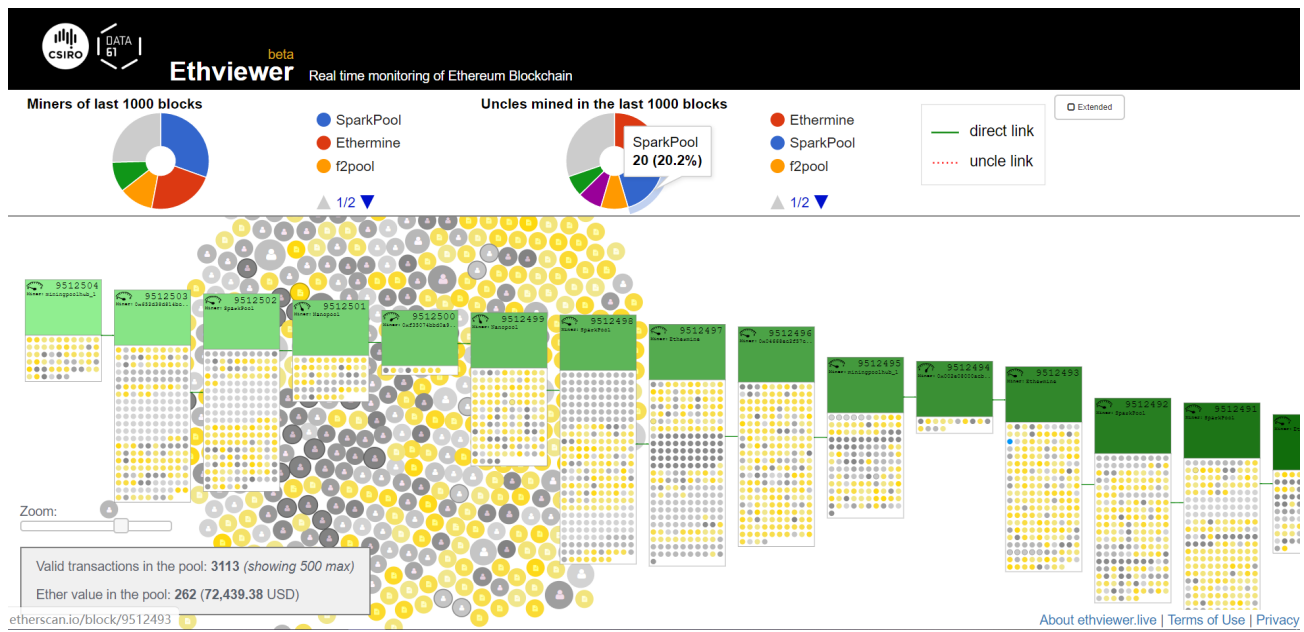

FIGURE 2: EXAMPLE FOR EXCITING VISUALIZATION. ETHEREUM BLOCKCHAIN (SOURCE: http://ethviewer.live/)

Activating learners around central course material is essential for the learning outcome and for the learner's ability to assess his or her learning progress. **The high number of students and assignments exacerbates the direct feedback by lecturers to evaluate or grade. The solution is to automate feedback and assessments using quizzes, discussion prompts, peer-feedback, and peer-graded assignments.**

### Feedback automation mechanisms
*Quizzes*
Having at least three to five quiz questions per 10 mins video lecture is recommended. Such quiz questions can also be related to, for example, articles or research in databases. **Quizzes are easy to automate, most often graded, and mandatory for completing the course. Questions can have pre-edited feedback and be presented to the learner immediately after a lecture or as a module test at the end of a module.**

Disclaimer
The creation of these resources has been (partially) funded by the ERASMUS+ grant program of the European Union under grant no. 2018-1-LT01-KA203-047044.
Neither the European Commission nor the project's national funding agency DAAD are responsible for the content or liable for any losses or damage resulting of the use of these resources.

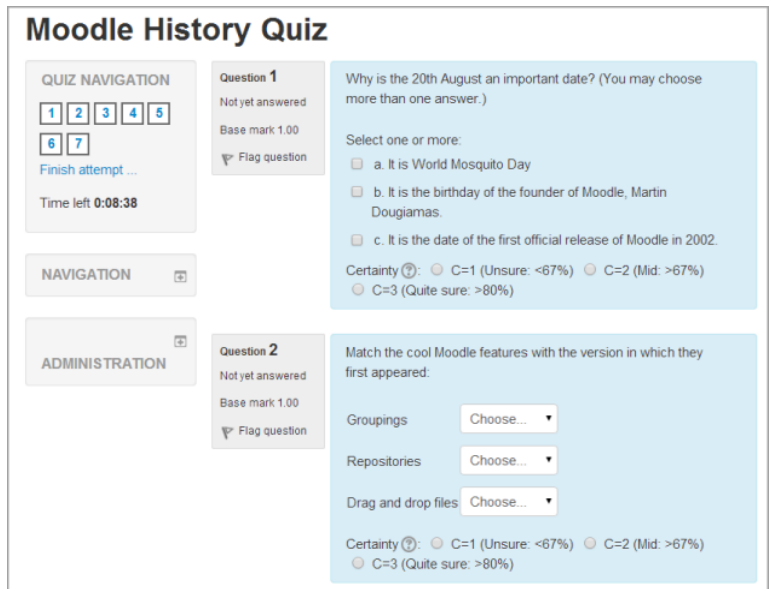

FIGURE 3: MOODLE QUIZ (SOURCE: MOODLE.ORG)

*Discussion Prompts*
A discussion prompt is an excellent technique for engaging a global audience of learners. Based on readings or video lectures, a discussion can be framed by stating a few good questions that can initiate a global peer-discussion.

*Peer-feedback and Peer-graded Assignments*
Peer-feedback and graded assignments let the learner engage more deeply with the content – and students learn a lot from evaluating each other's work. Here, it is crucial to describe well-defined review criteria.

*Use-case and Project Definition*
The curriculum is designed with a case study-based approach that employs case studies for teaching blockchain to interdisciplinary groups of students. The case studies need to be transformed into projects with a clear project scope (definition and goals) fostering the active learning of students. The projects should be developed to have different learning activities and material aligned to the course content. The material selection and activities vary but some suggestions for material types are:

- Detailed guideline for the case study or project work with concrete tasks and activities
- Presentation slides with background material (lecture, reading, statistics, use–case)
- Videos of lecture, e.g., 10-minute-long sequences, and an introduction trailer of the course
- Expert interviews (live, video, or podcast)
- Readings for self-studying
- Quizzes (assessment)
- Teaching case for group work (description and assessment methods)
- Boundaries of the case study or project including the objectives

The use-case and project descriptions should be elicited from stakeholders and at least comprise of the problem definition and context, utility of the problem solution, involved stakeholders (including references) and resources, and a clear goal definition. The usage of a template is advisable especially if multiple case studies or projects are used in the same course. A mapping of course learning goals and project activities are convenient to align course progress and project activities. The mapping can be used to estimate the project time necessary and adopting it. The project time should be at least half of the time allocated for learning activities of students to compensate for the upfront investment in introducing the case study or project. The preparation of the lecturer depends on her or his familiarity with the case study.

An effective case study analyses a real-life situation where existing problems are understandable by students and need to be solved. The theory should be related to a practical situation, e.g., applying the knowledge discussed in the course to a practical situation at hand in the case study. A case study should entail the following six activities:

1. Identify the problems in the case.
2. Select the main problems in the case.
3. Develop and suggest solutions to these main problems.
4. Analyze and evaluate solutions.
5. Recommend the best solution to be implemented.
6. Detail how this solution should be implemented.

### SCRIPT

Once a distinct target group and draft versions of the course outline, its learning objectives, quiz questions, resources, and central visuals are defined, scripting the course is necessary.

In general, making a video lecture is an excellent medium for engaging learners but very different from presenting a lecture in front of a class. A video does not provide feedback, and thus engaging an audience is more complicated. Additional preparation is necessary, and preparing a written manuscript containing guidelines for the video lecture creation might be helpful:

- The manuscript should be precise and well-structured.
- Graphs, illustrations, and other materials are clearly explained for the audience.
- Following the planned script during the recording helps to avoid making mistakes or getting unclear.
- Five hundred words for approximately four minutes of talk are adequate.
- Videos should be between 6-8 minutes for a focused learning session.

As stated before, this phase accommodates for adjusting the course outline and sharpen learning objectives in order to optimize the learning experience.

### RECORD AND EDIT

For recording the lecture, it is essential to work with as few interruptions by, e.g., noisy surroundings as possible. After recording, editing videos is the next step and is used to structure the video material before publishing the material. Here, a design guide can support a coherent layout and learning experience.

### REVIEW AND FINALIZE

When the video is edited, the lecturer should review it and take notes in a review document. The final editing is based on the review document.

### IMPLEMENT

Disclaimer
The creation of these resources has been (partially) funded by the ERASMUS+ grant program of the European Union under grant no. 2018-1-LT01-KA203-047044.
Neither the European Commission nor the project's national funding agency DAAD are responsible for the content or liable for any losses or damage resulting of the use of these resources.

The course needs to be implemented on a chosen platform. For this project Moodle Cloud platform has been chosen. The reasons for choosing Moodle Cloud are numerous:

1) most EU universities use Moodle;
2) Various EU and, in particular, ERASMUS+ projects developed material but more important in out project additional tools for Moodle;
3) Moodle Cloud is always featuring the newest version of Moodle which are backward compatible and thus eliminating need for transitioning course materials through different version updates of Moodle);
    1. Moodle Cloud is free of charge for 500 number of online users as of January 2020, which is a sufficient number for SNOC courses.
    2. Getting technical support is another important reason, as any online platform offers (too) many features.

The implementation of all course elements should be finalized at least one week before the course launch.

A challenge for SNOCs is the collaboration of participants. Participants of SNOCs are often geographically closer to each other and might have physical meetings. Participants of MOOCs, on the other hand, are unrelated and geographically distributed which makes physical meetings infeasible. Therefore, in the case of SNOC, online interaction of participants needs to be supported by tools, video chat, shared screens or design spaces, and maybe even by physical meeting venues. Part of the implementation is to prepare an online collaboration tool supporting joint learning activities as well as other interactions of the course participants.

### *PROMOTE*
When the course is open for pre-enrollment, the course responsible can begin the planned promoting activities.

### *LAUNCH*
During the first course session, the course responsible should monitor the learner's reactions closely. The learning platforms provide various data on how many learners have been reached, ratings, reviews, learner stories, reporting of issues, and questions in the course discussion forums.

### *OPTIMIZE*
The course regularly needs to be adjusted based on learner feedback. Direct feedback and suggestions of the learners can improve the course as well as adding new material.

## EDUCATIONAL APPROACHES FOR MOOCS AND SNOCS

### *BLENDED LEARNING*
Blended learning combines online education material and interaction with physical classroom methods. The purpose of blended learning is to promote the use of digital teaching methods, improve learning outcomes, and increase flexibility for staff and students[2]. Studies (Siemens, et al., 2015) suggest that the achievement of students was higher in blended learning environments. Blended learning might be an education approach allowing to extend or complement a SNOC because of its smaller number of learners. The material developed for a MOOC and SNOCs is usable in blended learning environments too.

---
[2] https://cobl.ku.dk/about/.

13Disclaimer
The creation of these resources has been (partially) funded by the ERASMUS+ grant program of the European Union under grant no. 2018-1-LT01-KA203-047044.
Neither the European Commission nor the project's national funding agency DAAD are responsible for the content or liable for any losses or damage resulting of the use of these resources.

## FLIPPED CLASSROOM

The flipped classroom signifies a particular format for blended learning where the one-way lecturing, or direct knowledge transfer from lecturer to a student, is moved from the classroom into preparatory online elements, thereby freeing up time in class for active learning and application of the knowledge attained in the online module. This approach is well-suited as a hybrid of classical teaching combined with online materials.

## INDIVIDUAL LEARNING PATHS

Personalized learning, learning in which the pace of learning and the instructional approach is optimized for individual learners, is facilitated using technology as well as suitable course design.

## E-MODERATING

Online learning needs a different moderating approach by teachers. E-moderating accompanies the course in close connection to the technical support as discussed in (Salmon, 2005) and depicted in Figure 4: 5 stage model from (Salmon, 2005)Salmon, 2005. From system access to development, teachers must support the educational development of students. The amount of interactivity increases from automatic tasks, e.g., providing access to the online platform by importing enrolled students from the university information system, up to development where individual response might be necessary to ensure the development of the students.

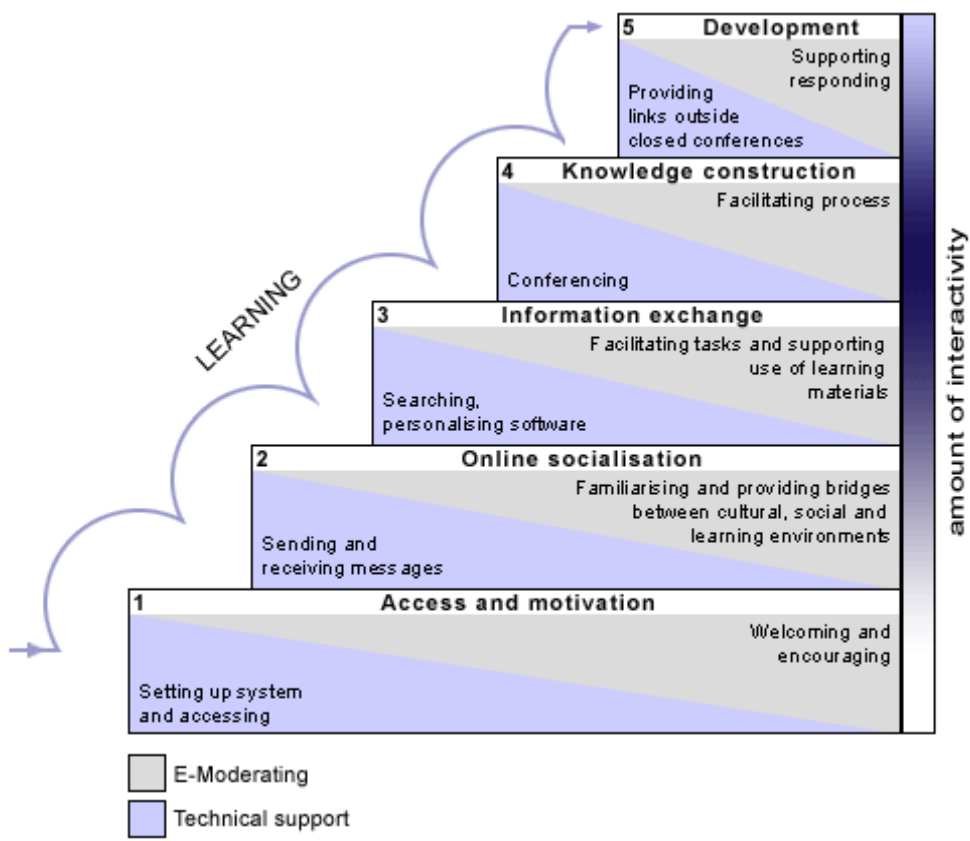

14Disclaimer
The creation of these resources has been (partially) funded by the ERASMUS+ grant program of the European Union under grant no. 2018-1-LT01-KA203-047044.
Neither the European Commission nor the project's national funding agency DAAD are responsible for the content or liable for any losses or damage resulting of the use of these resources.

FIGURE 4: 5 STAGE MODEL FROM (SALMON, 2005)

# DESIGN PRINCIPLES

The design principles applied in the curriculum design improve the learning outcome by a coherent design of individual modules.

## CONSTRUCTIVE ALIGNMENT

The student-centered teaching approach (Biggs, 1996) follows the shift from teaching to learning and from content-orientation to outcomes-orientation. It is a combination and variation of teaching and learning methods guaranteeing high-quality learning outcomes. Constructive alignment interweaves three core properties of teaching/learning:

- the underlying educational objective of the course,
- the practiced educational activity of the course, and
- the final assessment (exam).

This approach is more focused on students and their learning habits. The interaction between these core properties are: the interaction between teaching- and learning activities are aligned to educational objectives; exams assess the aspired educational objectives, and the teaching and learning has prepared students for these exams; all educational objectives are phrased in way to be testable on exams; all educational goals have to be reachable (in principle).

A teacher must ensure coherence between the core properties. The competency-oriented approach for developing exams must ensure capability to assess and evaluate the acquired student competencies against the intended learning goals and outcomes.

In order to achieve their learning goals, courses have to be organized around three categories of Teaching and Learning Activities (TLAs): lectures, exercise classes, and assignments. The TLAs are structured in a complementary fashion so that students are exposed to the learning goals in lectures, as well as take-home assignments and project work. Online exercises support assignments by developing skills and looking into the concepts; assignments allow students to develop and test their knowledge, skills, and competences. The evaluation for each course consists of a coarse quantitative and brief qualitative feedback of the assignments as well as a final exam, both targeted at assessing the achievement of the learning goals. We further investigate this basic triangle of *constructive alignment* - namely learning goals, TLAs, and evaluation - in the next sections.

Constructive alignment offers students to understand course objectives and guides their preparation (Eriksson, 2016) and (Cain & Babar 2016) as well as their retention (Hyyrö, 2017).

## EXPLICITLY FORMULATING LEARNING GOALS

The explicit formulation of teaching goals allows engaging in educational trade-offs consciously. Learning goals are integrated into teaching, for example, by:

Disclaimer
The creation of these resources has been (partially) funded by the ERASMUS+ grant program of the European Union under grant no. 2018-1-LT01-KA203-047044.
Neither the European Commission nor the project's national funding agency DAAD are responsible for the content or liable for any losses or damage resulting of the use of these resources.

- Dividing the lecture along with intended learning goals and define a minimal learning outcome of the lecture, which allows for more flexibility in teaching.
- Starting every lecture by defining the mutual learning contract, which includes the intended learning goals of the students.
- Referring to the connection to previous learning goals as a mental map of the course.
- Finishing the lecture by enumerating the same learning goals and link them to the lecture content,
- Phrasing the learning goals in a functional style. The purpose is to have explicit and assessable learning goals so that assessment activities can be aligned with the specific learning goals of the lectures. In particular, learning goals are phrased with active verbs, which reflect the Bloom's taxonomy (Bloom, 1994) in combination with Biggs's SOLO taxonomy (Chugh & Madhuravani, 2016) (Rodrigues & Santos, 2013) as a refined learning model applicable to online learning.

The students, as well as the lecturer, benefit from the explicit formulation of learning goals. The learning goals in the curriculum are based on the elicitation of skill demands from academia and industry partners in O1.2 of this project. The mapping of skills and learning goals was based on the course goal and the knowledge background of its participants. While the advantages to students are to be expected in the precise formulation of the teaching contract, the advantages of learning goals for a lecturer are more compelling. First, explicit learning goals allow for prioritization and, hence, increase the focus and effectiveness of lecture content and in-lecture exercises. Second, and more importantly, explicit learning goals are a tool for negotiating, choosing trade-offs, and justifying these choices. Skills are transformable in intended learning goals for interdisciplinary group projects. We followed in the design of the learning outcomes as well as the course descriptions the European guidelines presented in (Directorate-General for Education, 2015). The guidelines have been implemented by European universities for their educational programs.

## TEACHING METHODS

Each learning goal of a lecture is summarized by a critical viewpoint, forcing students to reflect on it. For example, a controversial viewpoint on current events related to the lecture topic engages students even though they might disagree. The exercises are designed based on the learning goals for the specific lecture; they also serve as a feedback channel to adapt the teaching modality and activate students.

Online and face-to-face lectures in this curriculum should be organized as follows:

- Checking of the learning environment, and removal of potential distractions;
- Presentation of learning goals;
- A recap of the previous section;
- An online quiz based on the active learning format;
- Declarative lecture for roughly half of one teaching segment (20 min);
- Controlled reflection phase for student activation
- A 3-5 minutes exercise for student activation and participation;
- An institutionalization phase potentially connected with an exposition of new concepts;
- A repetition of the format above for the second teaching segment, but excluding the recap and online quiz

Disclaimer
The creation of these resources has been (partially) funded by the ERASMUS+ grant program of the European Union under grant no. 2018-1-LT01-KA203-047044.
Neither the European Commission nor the project's national funding agency DAAD are responsible for the content or liable for any losses or damage resulting of the use of these resources.

- Multimedia sessions, i.e., videos have the highest learning impact with a 6-8 minutes length and introducing a final recall of the session's learning goals and presenting exemplary assessment criteria for achievement of these learning goals.

### ACTIVE LEARNING (WIEMAN)

In some learning activities -- such as the recapitulation of previous lecture content -- a clear structure for the discussion students should engage in is necessary. This structure should motivate discussions to move up in Bloom's objective taxonomy (Bloom, 1994), rather than turn into a simple recollection exercise. The approach of Wieman's active learning (Deslauriers, et al., 2011) is especially suited for teaching in large classes (Freeman, et al., 2014). For example, students must take a quiz at the beginning of the lecture for which they do not possess the right knowledge at this moment. *The motivation is that the student fails and learns during the lecture why his answer was wrong. It puts the student in an active position to learn why his answer was initially wrong*.

Other improvements can be directed to student activation and individual feedback in large student populations, for example, by using student-response-systems (clicker). The student-response-system allows for immediate feedback. A learning activity should be repeated if more than 50% of the students fail an in-lecture exercise.

It is important that the value and the goal of the activity are clear to the students. Here, a meaningful context to design a activity is supporting the student engagement, e.g., the context addresses a real-world issue, as well as a clear utility for the student in future. For example, in the context of the presented curriculum, a learning activity could be to retrieve and discuss news articles about blockchain security problems or implementations solving real-world problems. An introduction to interactive teaching using active learning with multiple examples can be found in (Barkley & Major, 2018).

### PROJECT-BASED LEARNING

Project-based teaching and learning is a dynamic teaching approach based on the belief that students acquire more profound knowledge through active experiences. This approach is very well suited for interdisciplinary learning activities (Crane, 2009), which is an inherent property of blockchain and the intention of this curriculum. We propose case study-based projects for interdisciplinary teams to facilitate knowledge dissemination and improving the overall learning outcome. A very good introduction with practical information on designing, preparing, and conducting project-based learning activities can be found in (Ho & Brooke, 2017).

### CASE-BASED TEACHING

The project- and case-based approach can be supported by including industry speakers into lectures presenting project cases. Additionally, these speakers should discuss project failures interactively with the students employing active learning activities (Freeman, et al., 2014).

## CURRICULUM

The modules in this curriculum are designed to deliver project-based teaching with online tools. The design follows the principles of constructive alignment. Course projects can be individual or group projects allowing

Disclaimer
The creation of these resources has been (partially) funded by the ERASMUS+ grant program of the European Union under grant no. 2018-1-LT01-KA203-047044.
Neither the European Commission nor the project's national funding agency DAAD are responsible for the content or liable for any losses or damage resulting of the use of these resources.

students to experience blockchain technology. The benefits of problem-based learning in online teaching have been studied in (An & Reigeluth, 2008). The curriculum is designed with one general, shared, cross-disciplinary module (Course 1a: Cross-Disciplinary) and four specific, single-discipline modules for cybersecurity, logistics and supply chain management, economy and management, and computer science.

## TIME VS. CREDIT POINTS

The conversion rate between learning hours and 1 ECTS point is between 25h to 30h of total workload of a student in the partner universities. The curriculum is designed with a calibrated hourly workload for the students according to this conversion rate. In the case of deviating conversion rates, hours need to be distributed accordingly.

## ASSESSMENT FORMS FOR MODULES AND RECOMMENDATIONS

### PRE-ASSESSMENT
A pre-assessment allows validating whether students meet the course pre-requisites. It can be used to guide the student's course enrollment and provides feedback to the course organizers.

### FORMATIVE ASSESSMENT
The formative assessment is given during the course and determines whether students are learning and how they react to the offered material. A formative assessment is linked to each course module and provides feedback to students on the learning progress. A formative assessment can be provided to the students before the exam to calibrate students to the exam level. Quizzes are well suited as formative assessment tools (Brindley, et al., 2009).

### SUMMATIVE ASSESSMENT
The summative assessment is used to measure students' achievement of the TLAs of the course. The assessment is essential part of the constructive alignment (Cain, Grundy und Woodward 2018). Approaches for digital assessment methods have been studied in (Lai & Sanusi, 2014).

## CURRICULUM INSTANTIATION
The curriculum is designed with one introduction course (Course 1A-C) to provide all students with the same level of knowledge, irrespective of their major. The courses are designed to allow to adapt to different credit requirements for courses in individual study programs. Course 1A is designed with 5 ECTS whereas courses 1B and 1C have 1 ECTS. It allows a course organizer to provide a course with 7 ECTS by combining the Course 1A with Course 1B and Course 1C. The course uses blended learning and project-based teaching to disseminate knowledge within the interdisciplinary project teams. Depending on the curriculum design goals, the course organizer instantiates one or multiple courses with discipline and field-specific content (Courses 2-5). The instantiation depicted in Figure 5[3] helps the course organizer ensure that the course follows the course design principles (cf. Design Principles) and teaching methods (cf. Teaching Methods). The syllabus contains references to learning material, e.g., project descriptions based on case studies. The course syllabus can be published and read by course participants. The final exam as summative assessment is responsibility of the course organizer and institution. The curriculum instantiation supports the autonomy of the course

---

[3] See the book (Wiggins & McTighe, 2005) for excellent suggestions from design to course instance.

Disclaimer
The creation of these resources has been (partially) funded by the ERASMUS+ grant program of the European Union under grant no. 2018-1-LT01-KA203-047044.
Neither the European Commission nor the project's national funding agency DAAD are responsible for the content or liable for any losses or damage resulting of the use of these resources.

organizer to decide on the exam form based on the regulation and hence a course organizer has to define the actual exam form. The recommended forms in the course description are summative assessments in oral or written form potentially as online assessments.

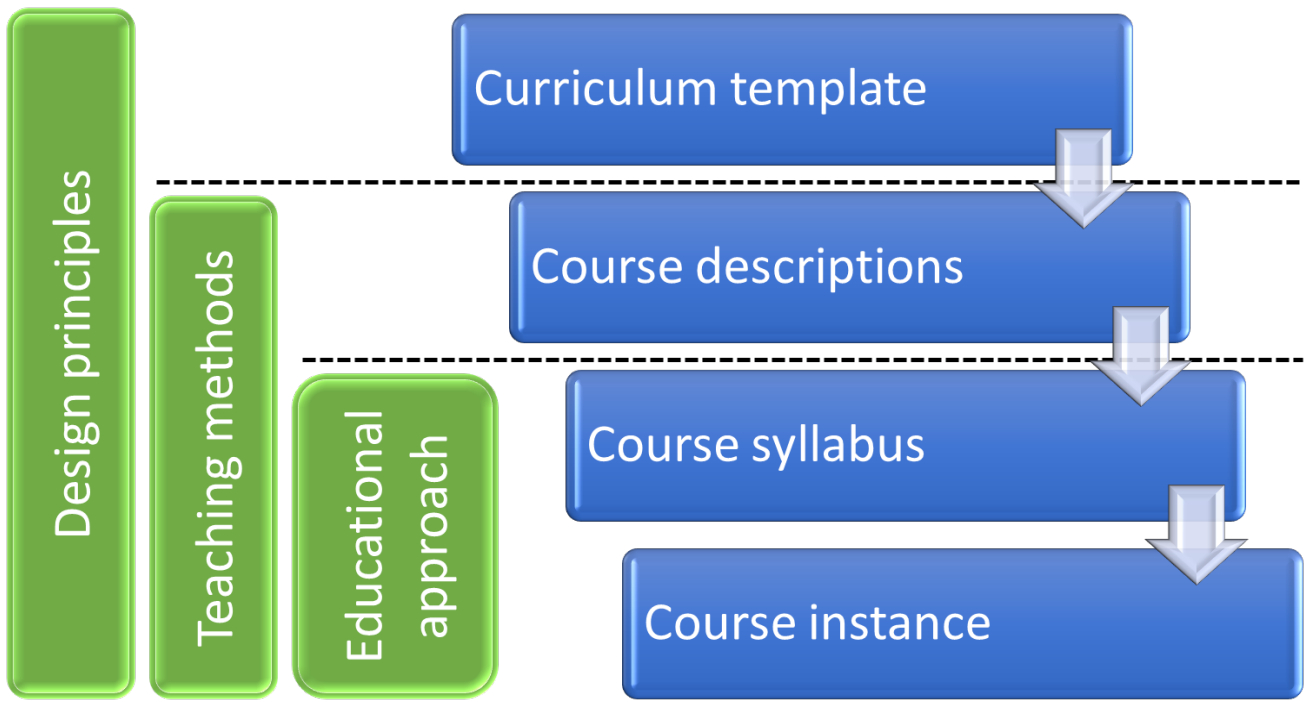

FIGURE 5: PROCESS FROM CURRICULUM TEMPLATE TO COURSE INSTANCE

*AIMS, OBJECTIVES AND LEARNING OUTCOMES*

This section sets out the overall aims of the course for master students and establishes that the course aims to create a learner-centered success culture, which will develop the graduates for Blockchain related interdisciplinary projects. The education goals are:

- Provide a stimulating and engaging learning experience and environment that offers cutting edge experience for students to develop their potential to become outstanding graduates and practitioners in an interdisciplinary setting.
- Educate graduates who have a foundational understanding of blockchain and its applications who can critically articulate their knowledge and able to demonstrate its applications with respect for others.
- Foster an academic community which promotes lifelong learning, supported by research, practice and problem based informed teaching and learning.
- Support flexible learning with technologies to reflect and anticipate student needs.

A series of Course Learning Outcomes were developed from the skills and competencies identified in the IO1 and are described in the BlockNet Competence Model (Fig. 4).

Disclaimer
The creation of these resources has been (partially) funded by the ERASMUS+ grant program of the European Union under grant no. 2018-1-LT01-KA203-047044.
Neither the European Commission nor the project's national funding agency DAAD are responsible for the content or liable for any losses or damage resulting of the use of these resources.

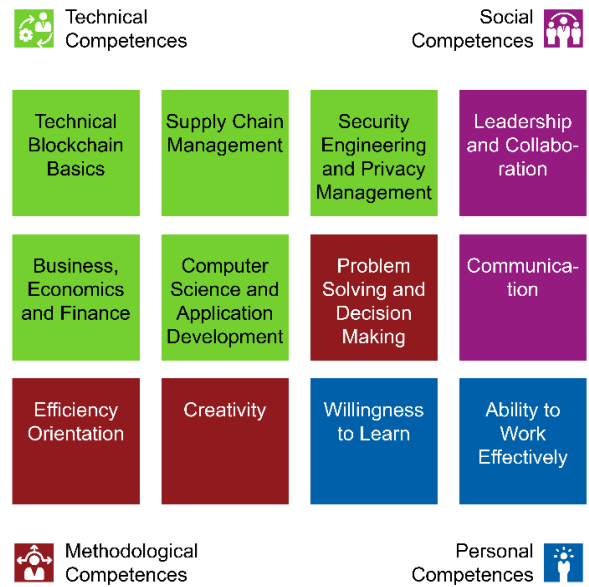

FIGURE 6: BLOCKNET COMPETENCE MODEL

These Course Learning Outcomes are in the left-hand column in (Table 1: Course Learning Outcomes / Subject knowledge and understanding Mapped to Modules) below and have been broken down into subject (specific) knowledge and understanding (i.e. Blockchain related interdisciplinary projects focus) and more general skills, which will be of use in multiple settings (including Blockchain related interdisciplinary projects). This table also shows the modules that these overall Course Learning Outcomes have been mapped to ensure that they are covered in the Course content.

| Course Learning Outcomes / Subject knowledge and understanding | Mapped to (modules) | Course Title | Course Type |
| --- | --- | --- | --- |

Disclaimer
The creation of these resources has been (partially) funded by the ERASMUS+ grant program of the European Union under grant no. 2018-1-LT01-KA203-047044.
Neither the European Commission nor the project's national funding agency DAAD are responsible for the content or liable for any losses or damage resulting of the use of these resources.

| | | | |
|---|---|---|---|
| **Identify and explain the foundations, architecture, concepts, principles, and technologies that were presented in the course, including their key terminology, underlying assumptions, and how they relate to one another.** | Introduction to distributed ledger technologies and properties (2h), business development (1h), and digital entrepreneurship (30 min) | Introduction in Blockchain and Applications | Course 1A: Cross-Disciplinary |
| **Compare the principal characteristics of blockchain platforms.** | Introduction to cryptography (2h) and information systems for blockchain (2h) and challenges | | |
| **Demonstrate how the theories, concepts, and technologies that were presented in the course were applied in the process of developing the prescribed blockchain project.** | Business process (2h) and requirement (2h) specification | | |
| **Exhibit the ability to apply blockchain technology through a written assignment that describes the blockchain innovation project and a tangible prototype that illustrates it.** | Cryptocurrencies (2h) and tokenization (1h) and token-based exchange mechanisms (1h) | | |
| **Evaluate the blockchain innovation project through theory-based critical assessment as well as discuss and present the results, including how to mitigate the prevailing challenges as well as how to move ahead with the development of the underlying project.** | Blockchain-enabled business process modeling (4h) | | |
| **To reflect the definition of interdisciplinarity and use methods of interdisciplinary cooperation such as IPBL.** | Introduction to platform and application development (1h), programming models (1h), and smart contract programming (2h) | | |
| **Evaluate the blockchain innovation project through theory-based critical assessment as well as discuss and present the results, including how to mitigate the prevailing challenges as well as how to move ahead with the development of the underlying project.** | Online case development and project presentations | | |
| **Identify branches, use-cases and concrete business processes to integrate and utilize blockchain-based solutions** | Blockchain-enabled business process modeling (4h) | Blockchain-enabled | Course 1B: Cross- |

Disclaimer
The creation of these resources has been (partially) funded by the ERASMUS+ grant program of the European Union under grant no. 2018-1-LT01-KA203-047044.
Neither the European Commission nor the project's national funding agency DAAD are responsible for the content or liable for any losses or damage resulting of the use of these resources.

| | | | |
|---|---|---|---|
| **Demonstrate how the theories, concepts, and technologies presented were applied in the identified business processes** | | Business Processes | Disciplinary |
| **Understand the effects of an integration for involved companies, departments, processes and employees** | | | |
| **Understand the most important blockchain frameworks, how they have evolved, and which programming languages they support to offer decentralized applications and smart contracts** | Introduction to platform and application development (1h), programming models (1h), and smart contract programming (2h) | Introduction in Blockchain Application Development | Course 1C: Cross-Disciplinary |
| **Apply basic programming models and smart contract programming** | | | |
| **Identify and explain the foundational theories, concepts, and technologies that were presented in the course, including their key terminology, underlying assumptions, and how they relate to one another.** | Introduction to cryptography (2h), distributed computing (1h), and information systems (1h) | Blockchain Foundations and Development | Course 2: Computer Science |
| **Exhibit the ability to apply blockchain technology through assessable code fragments.** | Blockchain components and systems development (1h), platforms (1h), and programming models (2h) | | |
| **Reflect on systems development and foundational challenges.** | | | |
| **Identify and explain the foundational theories, concepts, and technologies that were presented in the course, including their key terminology, underlying assumptions, and how they relate to one another.** | Logistics processes and information systems | Blockchain for Logistics and SCM | Course 3: Logistics and SCM |
| **Exhibit the ability to apply blockchain technology through assessable design documents.** | Blockchain components and security methods, business transformation, and blockchain-enabled logistics processes | | |
| **Reflect on systems design and foundational or specific challenges for logistics and supply chains.** | | | |
| **Identify and explain the foundational theories, concepts, and technologies that were presented in the course, including their key terminology, underlying assumptions, and how they relate to one another.** | Cryptographical methods and security design for distributed information systems | Blockchain for Enterprise IT Security | Course 4: Cyber Security |
| **Exhibit the ability to apply blockchain technology through assessable model fragments.** | Blockchain components and security methods, secure platforms and tools, and enterprise models | | |
| **Reflect on systems design and foundational or security challenges.** | | | |

| Identify and explain the foundational theories, concepts, and technologies that were presented in the course, including their key terminology, underlying assumptions, and how they relate to one another. | Business activities & processes and innovation management using information systems | Blockchain for Business | Course 5: Economy and Management |
|---|---|---|---|
| Exhibit the ability to apply blockchain technology through assessable design documents. | Blockchain components and security methods, business transformation, and blockchain-enabled business processes | | |
| Reflect on systems design and foundational or security challenges for business information systems. | | | |
| Improve personal and self skills. | existing modules at universities | Interpersonal Skills | Course 6: Pers. Skills |
| | | Pers. Development S. | Course 6: Pers. Skills |

TABLE 1: COURSE LEARNING OUTCOMES / SUBJECT KNOWLEDGE AND UNDERSTANDING MAPPED TO MODULES

*COMPETENCE MAPPING PROFESSIONAL COMPETENCES*

Description of Mapping for professional competences

| Technical Skills | | | |
|---|---|---|---|
| **Skill Cluster** | **Skill item** | **Bloom's Level** | **Link to curriculum** |
| **Technical BCT Basics** | Knowledge of the foundations, **general functionality, architecture,** components, principles (e.g., cryptocurrencies, wallets, smart contracts, separate platforms) of blockchain systems | 1 | Course 1: 1 |
| | Explain trust management principles | 2 | Course 1: 1 |
| | Define ways to maintain blockchain-based systems | 1 | |
| | Demonstrate blockchain technology capabilities and apply them to business-related challenges | 3 | Course 1: 4 |
| | Compare blockchain platforms to enable understanding of different system design choices | 2 | Course 1: 1,2 |
| | Discuss and compare different blockchain models, scheme and solutions with constructed/illustrated application (suggestions, proposals, methods for blockchain use in economics, business and finance) | 2 | |
| | **Comprehension of BCT application development** | 2 | Course 1: Module 6 |
| | **Knowledge of BC Use Cases and different application fields** | 1 | Course 1: 1 |
| | Knowledge of auditing, accounting and taxation processes as blockchain application fields | 1 | Course 5: Module 1 |

Disclaimer
The creation of these resources has been (partially) funded by the ERASMUS+ grant program of the European Union under grant no. 2018-1-LT01-KA203-047044.
Neither the European Commission nor the project's national funding agency DAAD are responsible for the content or liable for any losses or damage resulting of the use of these resources.

| | | | |
|---|---|---|---|
| **Business, Economics and Finance** | Knowledge of financial operations, sales, payments, and transactions impacted by blockchain solutions | 1 | Course 5: Module 2 |
| | Knowledge of regulatory standards, rules, laws, regulations, management standards relevant for blockchain implementations | 1 | Course 5: Module 2 |
| | **Comprehension of economic efficiency and ways to assess the profitability of BCT** | 2 | |
| | **Comprehension of market and customer needs in order to apply blockchain solutions** | 2 | |
| | **Ability to apply process designing methods (e.g. CMMN & BPMN)** | 3 | Course 1: Module 3 |
| | **Comprehension of Risk Management in BCT operations** | 2 | Course 1: Module 2 |
| **Supply Chain Management** | Knowledge of the general capabilities of blockchains in SCM | 1 | Course 3: Module 1 |
| | Explain the interoperability of blockchain technology and possible collaboration between unknown or untrusted parties in SCM | 2 | Course 3: Module 2 |
| | Demonstrate blockchain capabilities and apply them to counterfeit and fraud prevention problem statements | 3 | Course 3: Module 2 |
| | Demonstrate blockchain capabilities and apply them to provenance and track&trace problem statements in SCM | 3 | Course 3: Module 2 |
| | Analyze how information asymmetry in corporate networks can be addressed by the blockchain-based applications | 4 | Course 3: Module 2 |
| | **Comprehension of Supply Chain Management** | 2 | Course 3: Module 1 |
| | **Ability to design a concept for the use of BCT in SCM** | 5 | Course 3: Module 2 |
| **Computer Science and application development** | Comprehension of development processes for blockchain solutions | 2 | Course 2: Module 1 |
| | Knowledge of system architectures, frameworks, different layers | 1 | Course 2: 1 |
| | Discuss software quality goals and their impact on blockchain system development | 2 | |
| | Describe the software requirements elicitation and engineering process of blockchain systems | 2 | Course 1: Module 3 |
| | Develop and manage databases using data management systems (also use of SQL, etc.) | 5 | |
| | Knowledge of network protocols | 1 | |
| | Apply different programming languages **(e.g. P2P Programming)** | 3 | Course 2: Module 2 |
| | Develop a testing plan for concrete blockchain solutions | 5 | Course 2: Module 1 |
| | **Knowledge of Cryptocurrencies & Cryptocurrency Coding** | 1 | Course 1: Module 4 |
| | **Ability to apply methods of systems engineering** | 3 | |
| | **Comprehension of interface management** | 2 | |

Disclaimer
The creation of these resources has been (partially) funded by the ERASMUS+ grant program of the European Union under grant no. 2018-1-LT01-KA203-047044.
Neither the European Commission nor the project's national funding agency DAAD are responsible for the content or liable for any losses or damage resulting of the use of these resources.

|  | Ability to apply formal abstraction | 5 | Course 1: 3 |
|---|---|---|---|
|  | Knowledge of the connection between Data Science & BCT | 1 |  |
|  | **Abbility to programm smart contracts** | 5 | Course 2: Module 2 and Course 1: Module 6 |
| **Security Engineering and Privacy Management** | Describe privacy management principles using the blockchain solutions | 2 | Course 4: Module 2 |
|  | Explain identity management principles using the blockchain solutions | 2 | Course 4: Module 2 |
|  | Explain how data, information and processes can be secured by the use of the blockchain technology | 2 | Course 4: Module 2 |
|  | Recognise security countermeasure implications | 2 | Course 4: Module 1 |
|  | Explain access control (authentication, authorization and identity) models | 2 |  |
|  | Describe transaction protection and validation principles | 2 | Course 4: Module 1 |
|  | Underline major encryption and signature schemes | 2 | Course 4: Module 1 |
|  | State major fair mining principles | 2 |  |
|  | Identify security errors in smart contracts | 2 |  |

*SOCIAL AND PERSONAL COMPETENCES*

Description on how to integrate additional competences, e.g., social skills, than professional competences in our program.

| **Methodological Skills** | | | |
|---|---|---|---|
| **Skill Cluster** | **Skill item** | **Bloom's Level** | **Link to curriculum** |
| **Efficiency orientation** | Ability to transfer knowledge to internal (e.g., colleagues, developers, testers, etc.) and external (e.g., customers, support teams, and etc.) stakeholders | 6 | Course 1: 1 |
|  | Demonstrate ability to prioritise and to have a good time-management | 3 |  |
|  | Ability to organise interdisciplinary work | 5 | Course 1: 1 |
|  | **Ability to structure ones own work well** | 4 |  |
| **Creativity** | Sketch/imply creative solutions | 3 | Course 1: Module 1 & 3 |
|  | Apply innovative application development methods | 3 | Course 1: 5 |
|  | Practice creative solutions | 3 | Course 1: 5 |
|  | Knowledge of new interdisciplinary working methods | 1 |  |
| **Problem solving and Decision making** | Apply analytical methods to solve problems | 3 | Course 1: Module 3 |
|  | Apply evidence-based approaches for problem solving | 3 |  |
|  | Apply critical thinking | 3 |  |

Disclaimer
The creation of these resources has been (partially) funded by the ERASMUS+ grant program of the European Union under grant no. 2018-1-LT01-KA203-047044.
Neither the European Commission nor the project's national funding agency DAAD are responsible for the content or liable for any losses or damage resulting of the use of these resources.

| Social Skills | | | |
|---|---|---|---|
| **Leadership and Collaboration** | Lead and manage the team | 5 | Social skills will be conveyed via BlockNet Case Study. Advanced skills can be acquired by existing courses that we refer to in the next chapter. |
| | Demonstrate strong (inter-) organisational networking skills | 3 | |
| | Demonstrate ability to work in international and interdisciplinary teams | 3 | |
| | Practice to support colleagues with expert knowledge | 3 | |
| | Establish good social relationships with the customers | 3 | |
| | **Ability to allocate team members according to their specific competences in different BC tasks** | 3 | |
| | **Ability to work effective and collaborative in a team** | 3 | |
| | **Willingness to behave socially and ethically correct** | 3 | |
| | **Ability to mediate between different team members and align the knowledge level of different team members** | 6 | |
| **Communi-cation** | Demonstrate good written communication (documentation) skills | 3 | |
| | Demonstrate good verbal communication skills | 3 | |
| | Demonstrate ability to communicate complex and interdisciplinary problems | 3 | |
| | Use social media means | 3 | |
| | Demonstrate communication skills to internal and external stakeholders (colleagues, users, customers, advisors, and etc.) | 3 | |
| | Demonstrate good presentation skills | 3 | |
| | **Ability to communicate within intercultural teams** | 3 | |
| | **Ability to demonstrate negotiation skills considering different cultural background** | 3 | |
| Personal Skills | | | |
| **Willingness to learn** | Ability to learn quickly | 3 | Personal skills will be conveyed via BlockNet Case Study. Advanced skills can be acquired by existing courses that we refer to in the next chapter. |
| | Demonstrate Interest in new technology | 3 | |
| | Demonstrate Interest in continuing learning | 3 | |
| | Ability to accept and take into account feedback | 3 | |
| | Apply new ideas and be open-minded | 3 | |
| | **Ability to educate oneself** | 3 | |
| | **Willingness to work without existing partners/examples in the area of running BCT Projects** | 3 | |
| **Ability to work effectively** | Demonstrate ability to work independently and self-organized | 3 | |
| | Be proactive and take initiative | 3 | |
| | Be responsible, trusted, and committed | 3 | |
| | Ability to determine qualitative results | 3 | |
| | **Ability to act in a flexible manner and to adapt easily to new settings** | 3 | |
| | **Demonstrate honest and correct business practices** | 3 | |
| | **Demonstrate resilience and the ability to continue working on tasks despite difficulties** | 3 | |

Disclaimer
The creation of these resources has been (partially) funded by the ERASMUS+ grant program of the European Union under grant no. 2018-1-LT01-KA203-047044.
Neither the European Commission nor the project's national funding agency DAAD are responsible for the content or liable for any losses or damage resulting of the use of these resources.

EXISTING SOCIAL AND PERSONAL SKILLS COURSES TO BUILD UPON

In this section different existing courses on the topic of social and personal skills are presented to offer future BlockNet participants additional possibilities to acquire these skills on an advanced level. Through BlockNet case study group work social and personal skills will also be taught on a basic level.

INTERDISCIPLINARY COOPERATION

In order to operate successfully and across disciplines, additional competencies are required in addition to expertise in one's own field. Not only expertise in other fields, but also social skills and appreciative communication skills are decisive factors in the success of cooperation. It is time to break up "silo thinking", overcome departmental boundaries, put together mixed teams and promote interdisciplinary work. As the philosopher Carl Jaspers once said: "Truth begins with two". But what does interdisciplinarity really mean? These and other questions will be clarified in the training. In the course "important basics of interdisciplinarity" at TU Dortmund also practical topics will be discussed and directly tested in practical phases. In this way, students will be given a definition of the concept of interdisciplinarity and a distinction will be made, for example, between interdisciplinarity and multi-disciplinarity.

DIVERSITY QUALIFICATION

Diversity is one of the guiding concepts for the next generation of managers. Against the background of demographic change and the resulting shortage of skilled workers and globalization, diversity management is becoming an increasingly important task. In order to successfully advance a company, (globally) open, sensitized, highly qualified employees are sought today. Among other things, they must be able to deal productively with cultural and demographic diversity in their everyday work. For example, the workshop "Dealing with Diversity - Diversity Qualification for the Managers of Tomorrow" at TU Dortmund prepares students for future requirements in their jobs. It condenses exciting knowledge and practical application, for example on cultural particularities, explains how diversity management works, presents the diversity strategies of large companies, and shows the new directions being taken by recruiting in view of increasing diversity.

INTERCULTURAL COMMUNICATION

In the workshop "Intercultural Communication" at TU Dortmund, students can further develop their intercultural competences, one of the soft skills that is very much in demand in the profession. Since the seminar is held in English, they can also improve their communication skills in English.

WRITING AND PRESENTING

Technical articles, quality assurance documents, project reports, e-mails, project applications: Engineers often spend a large part of their working time on written documentation and communication. Thus, writing skills become an important key competence in the engineering profession, which can be successfully trained by every engineer. In addition, presentation skills are also crucial if you want to sell yourself and your products well and convey technical knowledge to a non-specialist audience in a skillful and interesting way. This competence is repeatedly demanded of engineers at conferences, during acquisitions or customer discussions, in meetings or in teaching. The workshop "Target Group Adaptive Writing and Presentations for Engineers - Basics" at TU Dortmund uses many exercises and individual consultation to convey in a compact and comprehensible way how professional success can be increased and communication can be made more effective with well thought-out writing.

27Disclaimer
The creation of these resources has been (partially) funded by the ERASMUS+ grant program of the European Union under grant no. 2018-1-LT01-KA203-047044.
Neither the European Commission nor the project's national funding agency DAAD are responsible for the content or liable for any losses or damage resulting of the use of these resources.

*BEING CREATIVE*

The "being creative" concept focus both on methodological skills about creativity, creativity techniques and the design of creative processes as well as person-oriented aspects for the development of one's own creative personality. In this way, students learn not only to develop creative ideas, but also to implement them in a self-confident and original way, so that ideas ultimately become innovations. Together with teachers from the engineering sciences, subject-specific teaching/learning scenarios will be designed to promote creativity, and workshops on creativity promotion will be offered to teachers and students alike. In view of the increasing importance of entrepreneurship, teaching/learning scenarios are also being developed in which students learn, to use their own creativity against the background of entrepreneurial thinking, use idea management to achieve innovations and to participate in the creation of an innovation-friendly organizational culture.

## COURSE STRUCTURE AND APPROACH TO TEACHING SECTION

This section sets out the structure of the Course and highlights the indicative nature of the modules and their interrelationships and individual institutions can make use of this in a menu-style manner (see Figure 7).

Disclaimer
The creation of these resources has been (partially) funded by the ERASMUS+ grant program of the European Union under grant no. 2018-1-LT01-KA203-047044.
Neither the European Commission nor the project's national funding agency DAAD are responsible for the content or liable for any losses or damage resulting of the use of these resources.

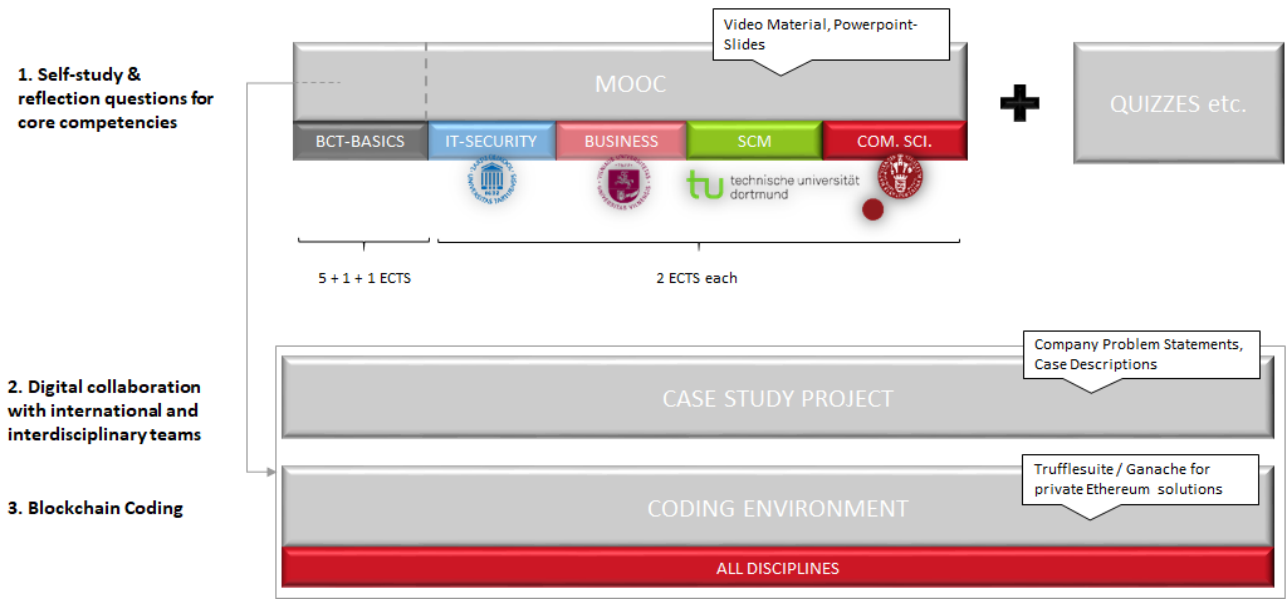

FIGURE 7. DESCRIPTION OF THE DESIGN AND COMPOSITION OF THE COURSE AND OF THE DIFFERENT MODULES, INCLUDING THE INTEGRATION OF THE CASE STUDY

Details of the individual courses and modules are shown in Appendix A, which contains a series of module descriptors, identifying the following for each module:

- Course title

- Course description

- Intended learning outcomes

- Indicative content

- Learning activities and teaching methods

- Assessment types

- Feedback to students

- Learning resources and key texts

In addition, there is a proposed live case study module shown in the Figures, in which students can develop a Blockchain related project and apply relevant concepts, theories, models and tools in a relevant organizational setting. As the requirements for this will vary significantly, an individual module descriptor has not been developed to cover this

Disclaimer
The creation of these resources has been (partially) funded by the ERASMUS+ grant program of the European Union under grant no. 2018-1-LT01-KA203-047044.
Neither the European Commission nor the project's national funding agency DAAD are responsible for the content or liable for any losses or damage resulting of the use of these resources.

# COURSE DESCRIPTIONS

## COURSE 1A: CROSS-DISCIPLINARY

*COURSE TITLE: INTRODUCTION IN BLOCKCHAIN AND APPLICATIONS*

ECTS:            5 ECTS

Language:        English

### COURSE DESCRIPTION
Blockchain has the potential to disrupt multiple industries by making transactions and processes more democratic, secure, transparent, and efficient. Building on this vision, students will explore how the capabilities and the underlying mechanism of blockchain can be applied to generate economic and social value. This course highlights the central topics of blockchain and distributed ledger technologies. It is designed to familiarize students with the foundations of blockchain technology and its key application areas. Practical case studies are presented by external industry stakeholders and interdisciplinary project teams will work on these case studies. Besides, the course seeks to provide hands-on experience in applying blockchain technology to developing novel insights and turning them into prototypes in self-organized project teams. In addition, the basics of interdisciplinary cooperation and methods such as Interdisciplinary Problem-based Learning (IPBL) are taught.

Intended learning outcomes by the end of the course, the students should be able to:

- Identify and explain the foundations, architecture, concepts, principles, and technologies that were presented in the course, including their key terminology, underlying assumptions, and how they relate to one another.
- Compare the principal characteristics of blockchain platforms.
- Demonstrate how the theories, concepts, and technologies that were presented in the course were applied in the process of developing the prescribed blockchain project.
- Exhibit the ability to apply blockchain technology through a written assignment that describes the blockchain innovation project and a tangible prototype that illustrates it.
- Evaluate the blockchain innovation project through theory-based critical assessment as well as discuss and present the results, including how to mitigate the prevailing challenges as well as how to move ahead with the development of the underlying project.
- To reflect the definition of interdisciplinarity and use methods of interdisciplinary cooperation such as IPBL.

### TEACHING METHODS
Teaching will be based on project-centric thematic online lectures (videos), digital exercises, studio work, case studies, field research, and online student presentations.

### EXAMINATION
Digital exam based on a product and ten standard pages written project. The product and the written project should be produced in parallel with the course.

### PREREQUISITES

Disclaimer
The creation of these resources has been (partially) funded by the ERASMUS+ grant program of the European Union under grant no. 2018-1-LT01-KA203-047044.
Neither the European Commission nor the project's national funding agency DAAD are responsible for the content or liable for any losses or damage resulting of the use of these resources.

None

## COURSE PLAN

| Module # | Topic |
|---|---|
| 1 | Introduction to distributed ledger technologies and properties (2h), business development (1h), and digital entrepreneurship (30 min) |
| 2 | Introduction to cryptography (2h) and information systems for blockchain (2h) and challenges |
| 3 | Business process (2h) and requirement (2h) specification |
| 4 | Cryptocurrencies (2h) and tokenization (1h) and token-based exchange mechanisms (1h) |
| 5 | Blockchain-enabled business process modeling (4h) |
| 6 | Introduction to platform and application development (1h), programming models (1h), and smart contract programming (2h) |
| 7 | Online case development and project presentations |

## STUDENT WORKLOAD

| Task | Total hours |
|---|---|
| Class lectures and exercises | 19 |
| Class preparation | 37 |
| Project preparation | 64 |
| Exam and exam preparation | 20 |
| **Total** | 140 |

**Software Tools and material provided by teachers**

## FEEDBACK TO STUDENTS

Students will receive feedback on their ongoing performance in a number of formal and informal ways, including:

Disclaimer
The creation of these resources has been (partially) funded by the ERASMUS+ grant program of the European Union under grant no. 2018-1-LT01-KA203-047044.
Neither the European Commission nor the project's national funding agency DAAD are responsible for the content or liable for any losses or damage resulting of the use of these resources.

Formative feedback will be made available within the delivery of the module, for example on student led tutorial activities, which will provide students the ongoing opportunity for their academic, personal and professional development and prepare them for the assessments.

Summative feedback will be clearly aligned with the assessment criteria, which will be made available with the assignment brief, and this will be given within an agreed deadline from the assessment submission date.

Disclaimer
The creation of these resources has been (partially) funded by the ERASMUS+ grant program of the European Union under grant no. 2018-1-LT01-KA203-047044.
Neither the European Commission nor the project's national funding agency DAAD are responsible for the content or liable for any losses or damage resulting of the use of these resources.

## COURSE 1B: CROSS-DISCIPLINARY

*COURSE TITLE: BLOCKCHAIN-ENABLED BUSINESS PROCESSES*

ECTS: 1 ECTS

Language: English

### INTENDED LEARNING OUTCOMES
By the end of the course, the students should be able to:

- Identify and explain the foundations, architecture, concepts, principles, and technologies that were presented in the course, including their key terminology, underlying assumptions, and how they relate to one another.
- Compare the principal characteristics of blockchain platforms.
- Demonstrate how the theories, concepts, and technologies that were presented in the course were applied in the process of developing a blockchain-enabled business model.
- Exhibit the ability to apply blockchain technology through a written assignment that describes the blockchain innovation project and a tangible prototype that illustrates it.
- Evaluate the blockchain innovation project through theory-based critical assessment as well as discuss and present the results, including how to mitigate the prevailing challenges as well as how to move ahead with the development of the underlying project.
- To reflect the definition of interdisciplinarity and use methods of interdisciplinary cooperation such as IPBL.

### TEACHING METHODS
Teaching will be based on project-centric thematic online lectures (videos), digital exercises, studio work, case studies, field research, and online student presentations.

### EXAMINATION
Digital exam based on a product and ten standard pages written project. The product and the written project should be produced in parallel with the course.

### PREREQUISITES
None

### COURSE PLAN

| Module # | Topic |
|---|---|
| 1 | Blockchain-enabled business process modeling (4h) |

### STUDENT WORKLOAD

| Task | Total hours |
|---|---|
| Class lectures and exercises | 4 |

Disclaimer
The creation of these resources has been (partially) funded by the ERASMUS+ grant program of the European Union under grant no. 2018-1-LT01-KA203-047044.
Neither the European Commission nor the project's national funding agency DAAD are responsible for the content or liable for any losses or damage resulting of the use of these resources.

| Class preparation | 7 |
| --- | --- |
| Project preparation | 13 |
| Exam and exam preparation | 4 |
| **Total** | 28 |

**Software Tools and material provided by teachers**

*FEEDBACK TO STUDENTS*

Students will receive feedback on their ongoing performance in a number of formal and informal ways, including:

Formative feedback will be made available within the delivery of the module, for example on student led tutorial activities, which will provide students the ongoing opportunity for their academic, personal and professional development and prepare them for the assessments.

Summative feedback will be clearly aligned with the assessment criteria, which will be made available with the assignment brief, and this will be given within an agreed deadline from the assessment submission date.

Disclaimer
The creation of these resources has been (partially) funded by the ERASMUS+ grant program of the European Union under grant no. 2018-1-LT01-KA203-047044.
Neither the European Commission nor the project's national funding agency DAAD are responsible for the content or liable for any losses or damage resulting of the use of these resources.

## COURSE 1C: CROSS-DISCIPLINARY

*COURSE TITLE: INTRODUCTION IN BLOCKCHAIN APPLICATION DEVELOPMENT*

ECTS: 1 ECTS

Language: English

### INTENDED LEARNING OUTCOMES

By the end of the course, the students should be able to:

- Identify and explain the foundations, architecture, concepts, principles, and technologies that were presented in the course, including their key terminology, underlying assumptions, and how they relate to one another.
- Compare the principal characteristics of blockchain platforms and development models.
- Demonstrate how the theories, concepts, and technologies that were presented in the course were applied in the process of developing the prescribed blockchain project.
- Exhibit the ability to apply blockchain technology through a written assignment that describes the blockchain innovation project and a tangible prototype that illustrates it.
- Evaluate the blockchain innovation project through theory-based critical assessment as well as discuss and present the results, including how to mitigate the prevailing challenges as well as how to move ahead with the development of the underlying project.
- To reflect the definition of interdisciplinarity and use methods of interdisciplinary cooperation such as IPBL.

### TEACHING METHODS

Teaching will be based on project-centric thematic online lectures (videos), digital exercises, studio work, case studies, field research, and online student presentations.

### EXAMINATION

Digital exam based on a product and ten standard pages written project. The product and the written project should be produced in parallel with the course.

### PREREQUISITES

None

### COURSE PLAN

| Module # | Topic |
|---|---|
| 1 | Introduction to platform and application development (1h), programming models (1h), and smart contract programming (2h) |

### STUDENT WORKLOAD

| Task | Total hours |
|---|---|
| | |

Disclaimer
The creation of these resources has been (partially) funded by the ERASMUS+ grant program of the European Union under grant no. 2018-1-LT01-KA203-047044.
Neither the European Commission nor the project's national funding agency DAAD are responsible for the content or liable for any losses or damage resulting of the use of these resources.

| Class lectures and exercises | 4 |
| --- | --- |
| Class preparation | 7 |
| Project preparation | 13 |
| Exam and exam preparation | 4 |
| **Total** | 28 |

**Software Tools and material provided by teachers**

*FEEDBACK TO STUDENTS*

Students will receive feedback on their ongoing performance in a number of formal and informal ways, including:

Formative feedback will be made available within the delivery of the module, for example on student led tutorial activities, which will provide students the ongoing opportunity for their academic, personal and professional development and prepare them for the assessments.

Summative feedback will be clearly aligned with the assessment criteria, which will be made available with the assignment brief, and this will be given within an agreed deadline from the assessment submission date.

Disclaimer
The creation of these resources has been (partially) funded by the ERASMUS+ grant program of the European Union under grant no. 2018-1-LT01-KA203-047044.
Neither the European Commission nor the project's national funding agency DAAD are responsible for the content or liable for any losses or damage resulting of the use of these resources.

# COURSE 2: COMPUTER SCIENCE

COURSE TITLE: BLOCKCHAIN FOUNDATIONS AND DEVELOPMENT

ECTS: 2 ECTS

Language: English

## INTENDED LEARNING OUTCOMES

By the end of the course, the students should be able to:

- Identify and explain the foundational theories, concepts, and technologies that were presented in the course, including their key terminology, underlying assumptions, and how they relate to one another.
- Exhibit the ability to apply blockchain technology through assessable code fragments.
- Reflect on systems development and foundational challenges.

## TEACHING METHODS

Teaching will be based on project-centric thematic lectures, digital exercises, and online feedback.

## EXAMINATION

Digital exam based on a digital test (quiz) and five standard pages written project. The written project should be produced in parallel with the course.

## PREREQUISITES

The student has successfully finished courses in discrete mathematics, software development, and information systems or equivalent courses.

## COURSE PLAN

| Module # | Topic |
| --- | --- |
| 1 | Introduction to cryptography (2h), distributed computing (1h), and information systems (1h) |
| 2 | Blockchain components and systems development (1h), platforms (1h), and programming models (2h) |

## STUDENT WORKLOAD

| Task | Total hours |
| --- | --- |
| Class lectures and exercises | 8 |
| Class preparation | 14 |
| Project preparation | 24 |

Disclaimer
The creation of these resources has been (partially) funded by the ERASMUS+ grant program of the European Union under grant no. 2018-1-LT01-KA203-047044.
Neither the European Commission nor the project's national funding agency DAAD are responsible for the content or liable for any losses or damage resulting of the use of these resources.

| Exam and exam preparation | 9 |
|---|---|
| **Total** | 55 |

*FEEDBACK TO STUDENTS*

Students will receive feedback on their ongoing performance in a number of formal and informal ways, including:

Formative feedback will be made available within the delivery of the module, for example on student led tutorial activities, which will provide students the ongoing opportunity for their academic, personal and professional development and prepare them for the assessments.

Summative feedback will be clearly aligned with the assessment criteria, which will be made available with the assignment brief, and this will be given within an agreed deadline from the assessment submission date.

Disclaimer
The creation of these resources has been (partially) funded by the ERASMUS+ grant program of the European Union under grant no. 2018-1-LT01-KA203-047044.
Neither the European Commission nor the project's national funding agency DAAD are responsible for the content or liable for any losses or damage resulting of the use of these resources.

## COURSE 3 (LOGISTICS AND SUPPLY CHAIN MANAGEMENT)

COURSE TITLE: BLOCKCHAIN FOR LOGISTICS AND SUPPLY CHAIN MANAGEMENT

ECTS: 2 ECTS

Language: English

### INTENDED LEARNING OUTCOMES
By the end of the course, the students should be able to:

- Identify and explain the foundational theories, concepts, and technologies that were presented in the course, including their key terminology, underlying assumptions, and how they relate to one another.
- Exhibit the ability to apply blockchain technology through assessable design documents.
- Reflect on systems design and foundational or specific challenges for logistics and supply chains.

### TEACHING METHODS
Teaching will be based on project-centric thematic lectures, digital exercises, and online feedback.

### EXAMINATION
Digital exam based on a digital test (quiz) and five standard pages written project. The written project should be produced in parallel with the course.

### PREREQUISITES
The student has successfully finished basic courses in logistics, supply chain management, or equivalent courses.

### COURSE PLAN

| Module # | Topic |
|---|---|
| 1 | Logistics processes (2h) and information systems (2h) |
| 2 | Blockchain components and security methods (2h), business transformation (1h), and blockchain-enabled logistics processes (1h) |

### STUDENT WORKLOAD

| Task | Total hours |
|---|---|
| Class lectures and exercises | 8 |
| Class preparation | 14 |
| Project preparation | 24 |

Disclaimer
The creation of these resources has been (partially) funded by the ERASMUS+ grant program of the European Union under grant no. 2018-1-LT01-KA203-047044.
Neither the European Commission nor the project's national funding agency DAAD are responsible for the content or liable for any losses or damage resulting of the use of these resources.

| Exam and exam preparation | 9 |
|---|---|
| **Total** | 55 |

*FEEDBACK TO STUDENTS*

Students will receive feedback on their ongoing performance in a number of formal and informal ways, including:

Formative feedback will be made available within the delivery of the module, for example on student led tutorial activities, which will provide students the ongoing opportunity for their academic, personal and professional development and prepare them for the assessments.

Summative feedback will be clearly aligned with the assessment criteria, which will be made available with the assignment brief, and this will be given within an agreed deadline from the assessment submission date.

## COURSE 4 (CYBER SECURITY)

*COURSE TITLE: BLOCKCHAIN FOR ENTERPRISE IT SECURITY*

ECTS: 2 ECTS

Language: English

### INTENDED LEARNING OUTCOMES

By the end of the course, the students should be able to:

- Identify and explain the foundational theories, concepts, and technologies that were presented in the course, including their key terminology, underlying assumptions, and how they relate to one another.
- Exhibit the ability to apply blockchain technology through assessable model fragments.
- Reflect on systems design and foundational or security challenges.

### TEACHING METHODS

Teaching will be based on project-centric thematic lectures, digital exercises, and online feedback.

### EXAMINATION

Digital exam based on a digital test (quiz) and five standard pages written project. The written project should be produced in parallel with the course.

### PREREQUISITES

The student has successfully finished courses in discrete mathematics, software development, and information systems or equivalent courses.

### COURSE PLAN

| Module # | Topic |
| --- | --- |
| 1 | Cryptographical methods (2h) and security design for distributed information systems (2h) |
| 2 | Blockchain components and security methods (2h), secure platforms and tools (1h), and enterprise models (1h) |

### STUDENT WORKLOAD

| Task | Total hours |
| --- | --- |
| Class lectures and exercises | 8 |
| Class preparation | 14 |

Disclaimer
The creation of these resources has been (partially) funded by the ERASMUS+ grant program of the European Union under grant no. 2018-1-LT01-KA203-047044.
Neither the European Commission nor the project's national funding agency DAAD are responsible for the content or liable for any losses or damage resulting of the use of these resources.

| Project preparation | 24 |
| Exam and exam preparation | 9 |
| **Total** | 55 |

*FEEDBACK TO STUDENTS*

Students will receive feedback on their ongoing performance in a number of formal and informal ways, including:

Formative feedback will be made available within the delivery of the module, for example on student led tutorial activities, which will provide students the ongoing opportunity for their academic, personal and professional development and prepare them for the assessments.

Summative feedback will be clearly aligned with the assessment criteria, which will be made available with the assignment brief, and this will be given within an agreed deadline from the assessment submission date.

Disclaimer
The creation of these resources has been (partially) funded by the ERASMUS+ grant program of the European Union under grant no. 2018-1-LT01-KA203-047044.
Neither the European Commission nor the project's national funding agency DAAD are responsible for the content or liable for any losses or damage resulting of the use of these resources.

## COURSE 5 (ECONOMY AND MANAGEMENT)

*COURSE TITLE: BLOCKCHAIN FOR BUSINESS*

ECTS:           2 ECTS

Language:       English

### INTENDED LEARNING OUTCOMES
By the end of the course, the students should be able to:

- Identify and explain the foundational theories, concepts, and technologies that were presented in the course, including their key terminology, underlying assumptions, and how they relate to one another.
- Exhibit the ability to apply blockchain technology through assessable design documents.
- Reflect on systems design and foundational or security challenges for business information systems.

### TEACHING METHODS
Teaching will be based on project-centric thematic lectures, digital exercises, and online feedback.

### EXAMINATION
Digital exam based on a digital test (quiz) and five standard pages written project. The written project should be produced in parallel with the course.

### PREREQUISITES
The student has successfully finished courses in business management and business innovation or equivalent courses.

### COURSE PLAN

| Module # | Topic |
|---|---|
| 1 | Business activities & processes and (2h) and innovation management using information systems (2h) |
| 2 | Blockchain components and security methods (2h), business transformation (1h), and blockchain-enabled business processes (1h) |

### STUDENT WORKLOAD

| Task | Total hours |
|---|---|
| Class lectures and exercises | 8 |
| Class preparation | 14 |

Disclaimer
The creation of these resources has been (partially) funded by the ERASMUS+ grant program of the European Union under grant no. 2018-1-LT01-KA203-047044.
Neither the European Commission nor the project's national funding agency DAAD are responsible for the content or liable for any losses or damage resulting of the use of these resources.

| Project preparation | 24 |

*FEEDBACK TO STUDENTS*

Students will receive feedback on their ongoing performance in several formal and informal ways, including:

Formative feedback will be made available within the delivery of the module, for example on student led tutorial activities, which will provide students the ongoing opportunity for their academic, personal and professional development and prepare them for the assessments.

Summative feedback will be clearly aligned with the assessment criteria, which will be made available with the assignment brief, and this will be given within an agreed deadline from the assessment submission date.

Disclaimer
The creation of these resources has been (partially) funded by the ERASMUS+ grant program of the European Union under grant no. 2018-1-LT01-KA203-047044.
Neither the European Commission nor the project's national funding agency DAAD are responsible for the content or liable for any losses or damage resulting of the use of these resources.

# EVALUATION

The curriculum will be evaluated in an interdisciplinary course organized by all project participants as part of a summer school. The evaluation of the course will be based on a participants' survey[4].

# CONCLUSION

The aim of this work was to develop didactical and organizational concept for interdisciplinary blockchain SNOC. The task-specific challenges were 1) to design a course for interdisciplinary audience of students of higher education institutions (HEI), and 2) to ground the theoretical content of the course with empirical examples and cases, thus anchoring the general theoretical concepts into specific knowledge/study fields of students. The produced curriculum concept defines the demands and requirements, standards, and conditions for the integration of blockchain SNOC in existing study programmes at the individual institutions.

Specifically, the following goals were set for this intellectual output:

1) This project designed a didactical and organizational concept for interdisciplinary Blockchain SNOC, facilitating remote learning opportunities leveraging educational access. We mainly define two different concepts of how courses can be delivered online. One is SNOC and another is MOOC. We differentiate SNOC from MOOC and describe the uniqueness of SNOC.

2) Different methods for teaching and learning for higher education study programmes are defined focussing on fostering collaborative work of international students from different institutions, respective study programs, and scientific disciplines, such as case-based teaching, project-based learning, active learning, and blended learning. The methods and their involvement have been outlined. Tailored online distance and active learning methods are applied.

3) We define a module structure, in which either oral or written exam can be held for each module, depending on the requirements and/or preferences of the implementing HEI. For different course, the modules can be instantiated to meet the needs of the individual institutional curriculum. A modular interdisciplinary course is used as an introduction and field-specific courses complement the interdisciplinary knowledge and skill set.

4) Recommendations for the exam form for each module has been presented and is adaptable to actual needs and requirements.

5) A series of course learning outcomes were developed from the skills and competencies according to European guidelines. Moreover, a series of course learning outcomes were developed from the skills and competencies according to European guidelines and based on the results from O1.

6) A review of methods to evaluate and measure prior knowledge and learning progress is discussed. Pre-assessment serves as a method to evaluate prior knowledge while formative and summative

---

[4] https://project-blocknet.eu.

Disclaimer
The creation of these resources has been (partially) funded by the ERASMUS+ grant program of the European Union under grant no. 2018-1-LT01-KA203-047044.
Neither the European Commission nor the project's national funding agency DAAD are responsible for the content or liable for any losses or damage resulting of the use of these resources.

assessment are used to measure learning outcome of different strategies. Moreover, the skills and BlockNet Competence Mosel also help to develop the course learning outcomes.

7) Develop and apply tailored online distance and active learning methods to develop explicit and tacit competences. It is pointed out that not only expertise, but also social skills and appreciative communication skills are decisive factors in the success of cooperation. Diverse qualification and intercultural communication help to develop the social and personal competence. Writing and presenting skills are also crucial. Moreover, subject-specific teaching/learning scenarios can be designed to promote creativity.

8) Demands, standards and conditions for integration of the SNOC in existing study programmes have been aligned to the European standard for ECTS. Thus, the organisational concept allows the integration of the SNOC in existing study programmes at the individual institutions. Demands, standards and conditions are covered.

Disclaimer
The creation of these resources has been (partially) funded by the ERASMUS+ grant program of the European Union under grant no. 2018-1-LT01-KA203-047044.
Neither the European Commission nor the project's national funding agency DAAD are responsible for the content or liable for any losses or damage resulting of the use of these resources.

Disclaimer
The creation of these resources has been (partially) funded by the ERASMUS+ grant program of the European Union under grant no. 2018-1-LT01-KA203-047044.
Neither the European Commission nor the project's national funding agency DAAD are responsible for the content or liable for any losses or damage resulting of the use of these resources.

Disclaimer
The creation of these resources has been (partially) funded by the ERASMUS+ grant program of the European Union under grant no. 2018-1-LT01-KA203-047044.
Neither the European Commission nor the project's national funding agency DAAD are responsible for the content or liable for any losses or damage resulting of the use of these resources.

Disclaimer
The creation of these resources has been (partially) funded by the ERASMUS+ grant program of the European Union under grant no. 2018-1-LT01-KA203-047044.
Neither the European Commission nor the project's national funding agency DAAD are responsible for the content or liable for any losses or damage resulting of the use of these resources.